\documentclass[aps,prl,twocolumn,showpacs]{revtex4}
\usepackage{graphicx,color}

\begin{document}

\title{Topology of Fracture Networks}

\author{Christian Andr{\'e} Andresen}
\email{c.a.andresen@gmail.com}
\author{Alex Hansen}
\email{Alex.Hansen@ntnu.no}
\affiliation{Department of Physics, Norwegian University of Science and
Technology, N--7491 Trondheim, Norway}
\author{Romain Le Goc}
\email{r.legoc@itasca.fr}
\author{Philippe Davy}
\email{philippe.davy@univ-rennes1.fr}
\affiliation{Geosciences Rennes, UMR 6118, CNRS, Universit{\'e} 
de Rennes 1, F-35042 Rennes, France}
\author{Sigmund Mongstad Hope}
\email{hope@polytec.no}
\affiliation{Department of Physics, Norwegian University of Science and
Technology, N--7491 Trondheim, Norway}
\affiliation{Polytec R \& D Institute, S{\o}rhauggata 128,
N--5527 Haugesund, Norway}

\date{\today}

\begin{abstract}
We propose a mapping from fracture systems consisting of intersecting fracture
sheets in three dimensions to an abstract network consisting of nodes and 
links.  This makes it possible to analyze fracture systems with the methods 
developed within modern network theory.  We test the mapping for 
two-dimensional geological fracture outcrops and find that
the equivalent networks are small-world and dissasortative.  By
anlayzing the Discrete Fracture Network model, which is used to generate  
artifical fracture outcrop networks, we also find small world networks. However, the networks turn out to be assortative.  
\end{abstract}

\pacs{89.75.Hc,91.55.Jk,81.40.Np,89.75.Kd}
\maketitle
Topological analysis of networks has had an explosive growth over the last
decade \cite{b03}. A large number of new concepts and quantitive tools 
for describing networks have been introduced, making it possible to describe 
and classify complex network structures at a level that never earlier has 
been achieved \cite{ab02,blmch06}.
There is one class, though, of networks that has resisted this kind of 
analysis: Fracture networks.  These consist of intersecting fracture 
{\it sheets,\/}
making both the concepts of links and nodes far from obvious.  Fracture
networks, however, are extremely important from a technological point of
view.  For example, in carbonate petroleum reservoirs, the oil is transported
through fracture networks as the permeability of the porous matrix is too
low \cite{g07}.  Another example is the extraction of shale gas though 
hydrofracturing \cite{m11}.   

We propose a transformation from fracture network to an equivalent network consisting of nodes and links.  This makes it possible to qualitatively and quantitatively characterize the topology of fracture networks.

An important consequence of this is that it is possible to compare models that generate artificial networks with real networks quantitatively.  

Fracture {\it outcrop\/} networks have been studied from a network point of 
view by Valentini et al.\ \cite{vpp07a,vpp07b}.  Fracture outcrops are 
fracture lines visible on the surfaces of geological formations.
The outcrop fracture lines are one-dimensional cuts through the 
two-dimensional fracture sheets. Valentini et al. treats fracture lines as links and their crossing points as nodes, this gives a more narrow degree distribution than the transform proposed in this paper. However, Valentini et al. also conclude that fracture networks are small-world networks\cite{vpp07a}. In three dimensions where the fractures are sheets, the transformation we propose is
necessary to define the topology network.  

Our analysis is somewhat related to the information measure for cities introduced by Rosvall et al.\ \cite{rtms05}.

\begin{figure}
\includegraphics[scale=0.45,clip]{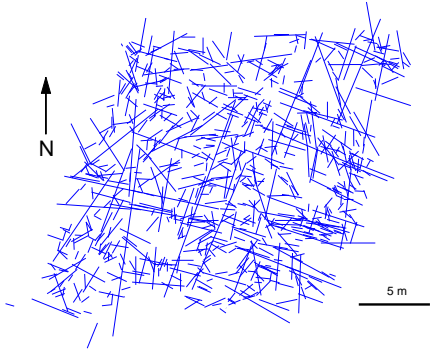}
\includegraphics[scale=0.25,clip]{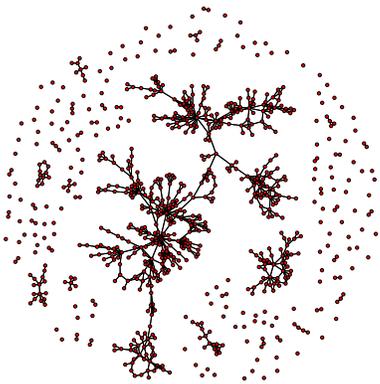}
\caption{\label{fig1} (Color online) a) Fracture network of outcrop AMS000025 
b) Equivalent network based on the original network shown in a).
} 
\end{figure}

Lacking data on three-dimensional fracture systems, we 
analyse in the following fracture data from eight outcrops found in
south-east Sweden. A detailed description of the bedrock 
composition and geological history are given in 
\citep{sasw08,ddbd06}.
We show one of the outcrop fracture networks in Fig.\ \ref{fig1}a.  
As we shall see, the equivalent network (shown in Fig.\ \ref{fig1}b) 
constructed from the original network has small-world character. 
Furthermore, it is {\it disassortative.\/}  

We then go on to analyse artificial fracture networks generated with the
Discrete Fracture Network (DFN) model \cite{dbdd03}.  The equivalent networks 
constructed from the original networks generated by this model also show small-world behavior.  However, they are {\it assortative.\/} 

\begin{figure}
\includegraphics[scale=0.25,clip]{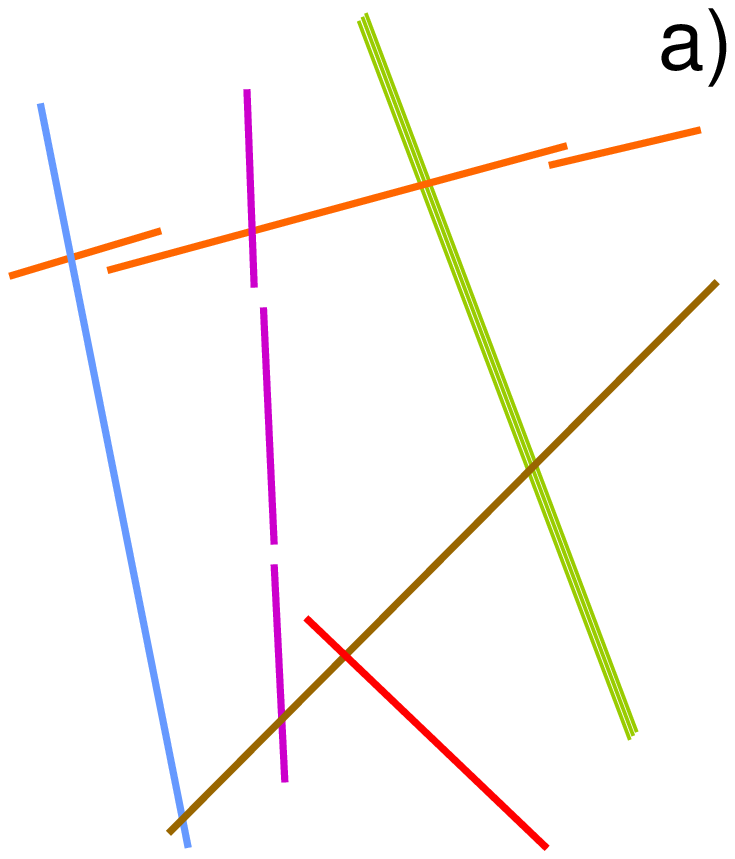}
\includegraphics[scale=0.25,clip]{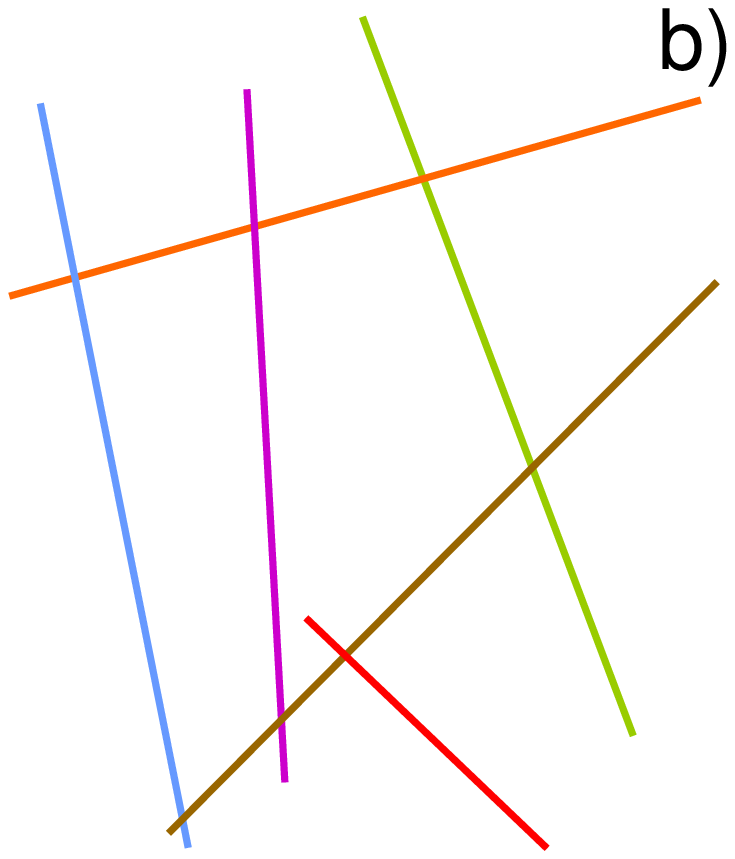}
\includegraphics[scale=0.25,clip]{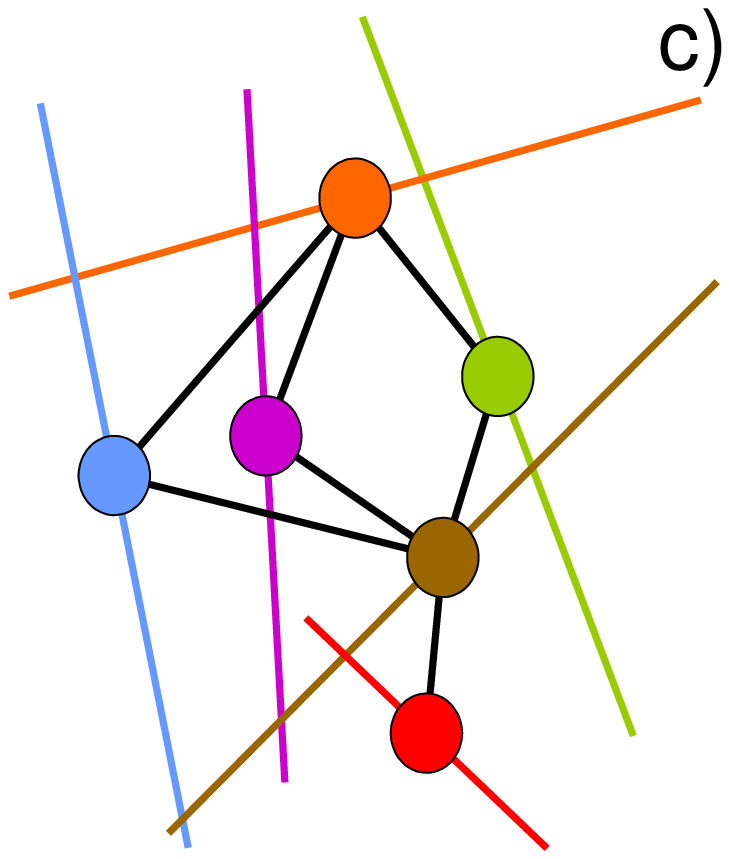}
\includegraphics[scale=0.25,clip]{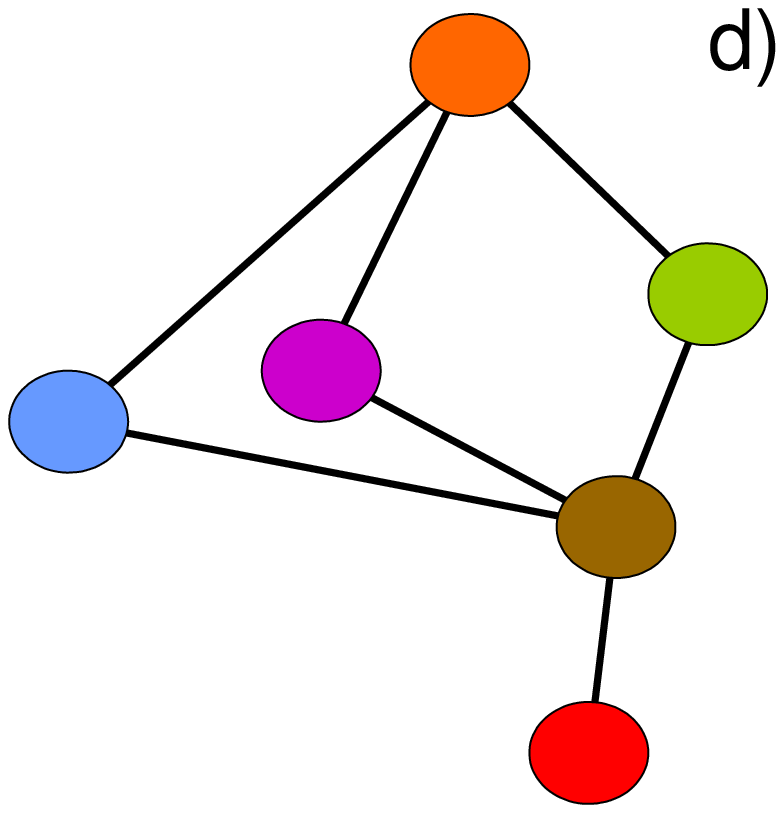}
\caption{\label{fig2} (Color online) Clock-wise from upper left. 
a) Representation of fracture outcrop network.
b) Reconnected fracture network. c) Equivalent network
placed on top of fracture outcrop network. d) Equivalent network 
representation of b).
} 
\end{figure}

The eight outcrops covers between 250
and 600 $\textrm{m}^2$. All visible fractures with length over 0.5 m have been
recorded in the data sets.  We prepare the data sets as follows. When tracing 
the fracture lines, they may appear disconnected or doubled due to topography
or ground weathering.  An illustration of a outcrop is shown in Fig.\ \ref{fig2}a. We
therefore use a reconnection procedure \cite{ddgdb09}. That is, we first 
project fracture traces on a flat surface to reduce the perturbation due to 
rock surface topography. Then scattered segments that are likely to belong to 
the same trace are reconnected to one single segment accounting for orientation 
and distance consistency. We focus on traces with a dashed-line, 
disconnected step or layered patterns. We then straighten all the fractures 
lines. The result is shown in Fig.\ \ref{fig2}b.

\begin{figure}
\includegraphics[scale=0.35,clip]{fig3a.eps}
\includegraphics[scale=0.4,clip]{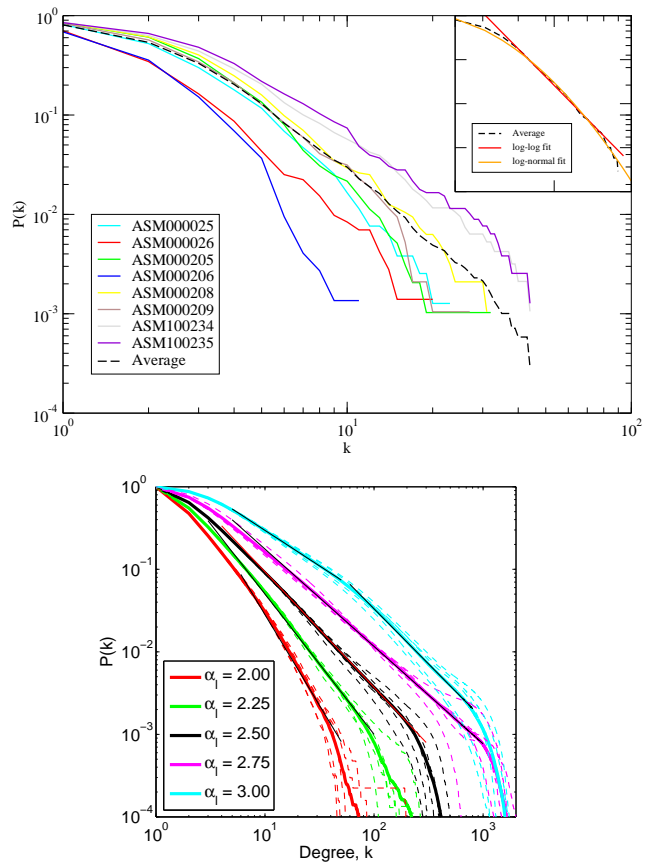}
\caption{\label{fig3} (Color online) 
Cumulative degree distribution for: a) Networks generated from the eight outcrop data sets.  Insert shows the average compared to a log-log (exponent $-2.3$) and a log-normal fit. b) DFN model. Values for the expoents of the fits are given in Table \ref{table2}.
} 
\end{figure}

We have now come to the central idea of this paper.  In Fig.\ \ref{fig2}c,
each fracture line has been associated with a node.  {\it Whenever two 
fracture lines  cross, we place a link between the nodes 
representing the two fracture lines.\/}  In Fig.\ \ref{fig2}d, 
we show the equivalent network consisting of nodes representing 
the fracture lines and links representing 
crossing fracture lines \cite{a08}. 

We note that this equivalent network is as simple to 
construct in a {\it three-dimensional system of fracture 
sheets:\/} each fracture sheet is represented by a node and whenever 
two sheets cross each other, a link is placed between the equivalent nodes.  

Arguably the most central property of any complex network is the degree 
distribution $p(k)$. The degree, $k$, of a node is the number of other nodes 
that it is linked to. The equivalent networks generated from the outcrop
networks show a broad degree distribution. We plot the cumulative distribution, 
$P(k)$ in Fig.\ \ref{fig3}. When $P(k)$ follows a power-law the network is scale free \cite{ab02}. We plot the data against log-log fits, but note that our geological data can be well fited by a log-normal. In the case of the DFN data the goodness of the log-log fit is dependent on model parameters. 

The clustering is a local measure of how well a network is connected on a 
local neighbor-to-neighbor scale. The global clustering coefficient, $C$, 
is defined \cite{ws98,n03} as the average over all the local
clustering coefficients, $C_i$, for each node
\begin{equation}
\label{clust}
C = \frac{1}{N} \sum_{i=1}^{i=N} C_i = \frac{1}{N} \sum_{i=1}^{i=N}
\frac{2E_{NN,i}}{k_i(k_i-1)},
\end{equation}
where $k_i$ is the degree of node $i$, $N$ is the total number of nodes and 
$E_{NN,i}$ is the number of links between the nearest neighbors of node $i$. 
The clustering coefficient falls in the interval $0 \le C \le 1$, and a high 
value indicates that there is a high chance that two neighbors of a node is 
connected to each other. This makes the network highly connected on a local 
scale, making it easy for nodes to efficiently interact on this scale.

\begin{table*}
\caption{List of the number of nodes (fractures),
links, maximum degree $k_{max}$, average
degree $\bar{k}$, clustering coefficient $C$,
clustering coefficient for rewired networks
$C_{RW}$, clustering coefficient for random
networks $C_{RA}$, efficiency $E$,
efficiency for rewired networks $E_{RW}$,
and efficiency  for random networks $E_{RA}$ for all the outcrop samples.}
\begin{tabular}{l c c c c c c c c c c}
\hline
\hline
Sample     &  Nodes & Links & $k_{max}$ & $\bar{k}$ & $C$
& $C_{RW}$ &  $C_{RA}$  &   $ E$  & $E_{RW}$ & $E_{RA}$\\
\hline
AMS000025  &  787   & 858   &     23   &
2.18   & 0.170  &  0.0048  &  0.00178  &  0.046  & 0.104   &  0.101 \\
AMS000026  &  716   & 520   &     20   &
1.45   & 0.088  &  0.0033  &  0.00087  &  0.019  & 0.048   &  0.032 \\
AMS000205  &  973   & 1188  &     32   &
2.44   & 0.193  &  0.0043  &  0.00174  &  0.032  & 0.122   &  0.118 \\
AMS000206  &  737   & 487   &     11   &
1.32   & 0.120  &  0.0013  &  0.00067  &  0.004  & 0.033   &  0.020 \\
AMS000208  &  955   & 1297  &     31   &
2.72   & 0.226  &  0.0067  &  0.00213  &  0.079  & 0.138   &  0.138 \\
AMS000209  &  955   & 1162  &     27   &
2.43   & 0.177  &  0.0050  &  0.00178  &  0.068  & 0.119   &  0.118 \\
AMS100234  &  946   & 1549  &     44   &
3.27   & 0.236  &  0.0138  &  0.00291  &  0.133  & 0.164   &  0.172 \\
AMS100235  &  785   & 1392  &     44   &
3.55   & 0.243  &  0.0180  &  0.00394  &  0.141  & 0.176   &  0.192 \\
\hline
Average    &  857   & 1057  &     29   &
2.42   & 0.182  &  0.0072  &  0.00198  &  0.065  & 0.113   &  0.111 \\
\hline
\hline
\end{tabular}
\label{table1}
\end{table*}

In order to determine whether the clustering coefficients found for the 
networks are large for their number of nodes and links, we compare them to 
rewired and random versions of the same networks.
In rewiring \cite{mg01} two pairs of connected nodes are selected at random, 
and the links interchanged so that two new pairs of connected nodes are 
created. The procedure is repeated until all links are moved. This preserves 
the degree distribution since all nodes retain their initial degree, but it 
removes any correlation between the degrees of the connected nodes. For the 
random version all links are removed and redistributed randomly between the 
nodes. This produces a new degree distribution that is generally not broad. 
In all cases the quoted values for these networks are averaged over 1000 
realizations.  As can be seen from Table \ref{table1}, the equivalent networks 
have an average clustering coefficient of 0.18 which is more than an order of 
magnitude larger than for comparable rewired networks, and two orders of 
magnitude larger than for purely random versions. Hence, they are well 
connected on a local scale.

The efficiency, $E$, is a global measure for how well the different parts of 
the network are connected, and how easily nodes in different parts of the 
network can interact. The measure is defined using the shortest distance, 
$d_{ij}$, between two nodes $i$ and $j$ \cite{blmch06}
\begin{equation}
\label{eff}
E = \frac{1}{N(N-1)} \sum_{(i,j) \in N, i \neq j} \frac{1}{d_{ij}},
\end{equation}
where $d_{ij} = \infty$ if node $i$ and $j$ are not connected. $E$ falls in 
the interval $0 \leq E \leq 1$, and a high value indicates that it is easy for 
nodes far apart in the network to interact since there on average is just a 
few links between any two nodes.

In Table \ref{table1} we present $E$ for all the equivalent networks and their 
average is 0.065, which is smaller than for the rewired ($E_{RW}$) and 
random ($E_{RA}$) versions both having an average of 0.11. However the
efficiency ($E$) is only smaller by a factor of about 2, 
making $E$ and $E_{RW/RA}$ of the same order. We would expect the rewired and 
random networks to have a high efficiency, several orders of magnitude larger 
than ordered networks, because they have a large portion of long-range links. 
The fact that the equivalent networks have an efficiency comparable to that of 
the rewired and random versions means that compared to ordered networks they 
have a large efficiency. We will discuss the impact of $C$ and $E$ for 
the equivalent networks in more detail below.

\begin{figure}
\includegraphics[scale=0.35,clip]{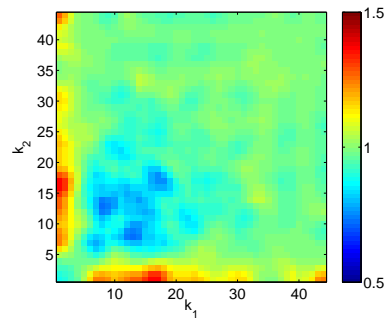}
\caption{\label{fig4} (Color online) Plot of the correlation matrix
$C(k_1,k_2)$ based on the equivalent networks generated from the eigth
outcrop fracture data sets.
} 
\end{figure}

It is also interesting to study any correlations between the degrees of linked
nodes. Does high degree nodes link predominantly to low degree nodes or high 
degree nodes? Maslov and Sneppen \citep{ms02} introduced a correlation matrix
\begin{equation}
\label{ms}
C(k_1,k_2) = \frac{P(k_1,k_2)}{P_R(k_1,k_2)},
\end{equation}
where $P(k_1,k_2)$ is the probability that a node of degree $k_1$ is 
linked to a node of degree $k_2$ for the network to be investigated. 
$P_R(k_1,k_2)$ is the same probability of a rewired version of the network. 
If $C(k_1,k_2) = 1$ for all $(k_1,k_2)$ then there is no degree correlations 
in the linking between nodes. If $C(k_1,k_2) > 1$ for some values of 
$(k_1,k_2)$ then there is an over-representation of links between nodes of 
degree $k_1$ and $k_2$ in the investigated network compared to that of a 
rewired version of the network. If $C(k_1,k_2) < 1$ there is an 
under-representation. Note that the matrix $C(k_1,k_2)$ is symmetric.

In Fig.\ \ref{fig4} we have plotted the average of the matrix $C(k_1,k_2)$ 
for all outcrops, where $P_R(k_1,k_2)$ is averaged over 10000 realizations. 
We observe an over-representation of small degree nodes linking to higher 
degree nodes, and an under-representation of equal degree nodes linking to 
each other. Such networks are disassortative, and are abundant in naturally 
occurring networks \citep{rt04,hh07}.

The characteristic path length, $L$ is defined as the average distance 
between any pair of nodes of a network,
\begin{equation}
\label{length}
L = \frac{1}{N(N-1)} \sum_{(i,j) \in \textit{N}, i \neq j} d_{ij}.
\end{equation}
Having a large clustering coefficient indicates a large local connectivity, 
and a small characteristic path length indicates a large global 
connectivity. When both of these criteria are fulfilled, we have a 
small-world network \cite{ws98}. Networks consisting of more than one 
disjoint part will have $d_{ij}=\infty$
for at least one pair of nodes. Hence, the characteristic path length 
is not a good measure for the global connectivity of such networks. However a 
small value of $d_{ij}$ for most pairs of nodes will give a large average 
value for $1/d_{ij}$ which is measured by the efficiency. Therefore a large 
$E$ is comparable to a small $L$ for describing the global connectedness. 
Since the fracture networks found in the outcrops have been shown to have a 
clustering coefficient significantly larger than rewired and random versions, 
and an efficiency of the same order as the rewired and random networks we 
conclude that these are small-world networks.

\begin{figure}
\includegraphics[scale=0.2,clip]{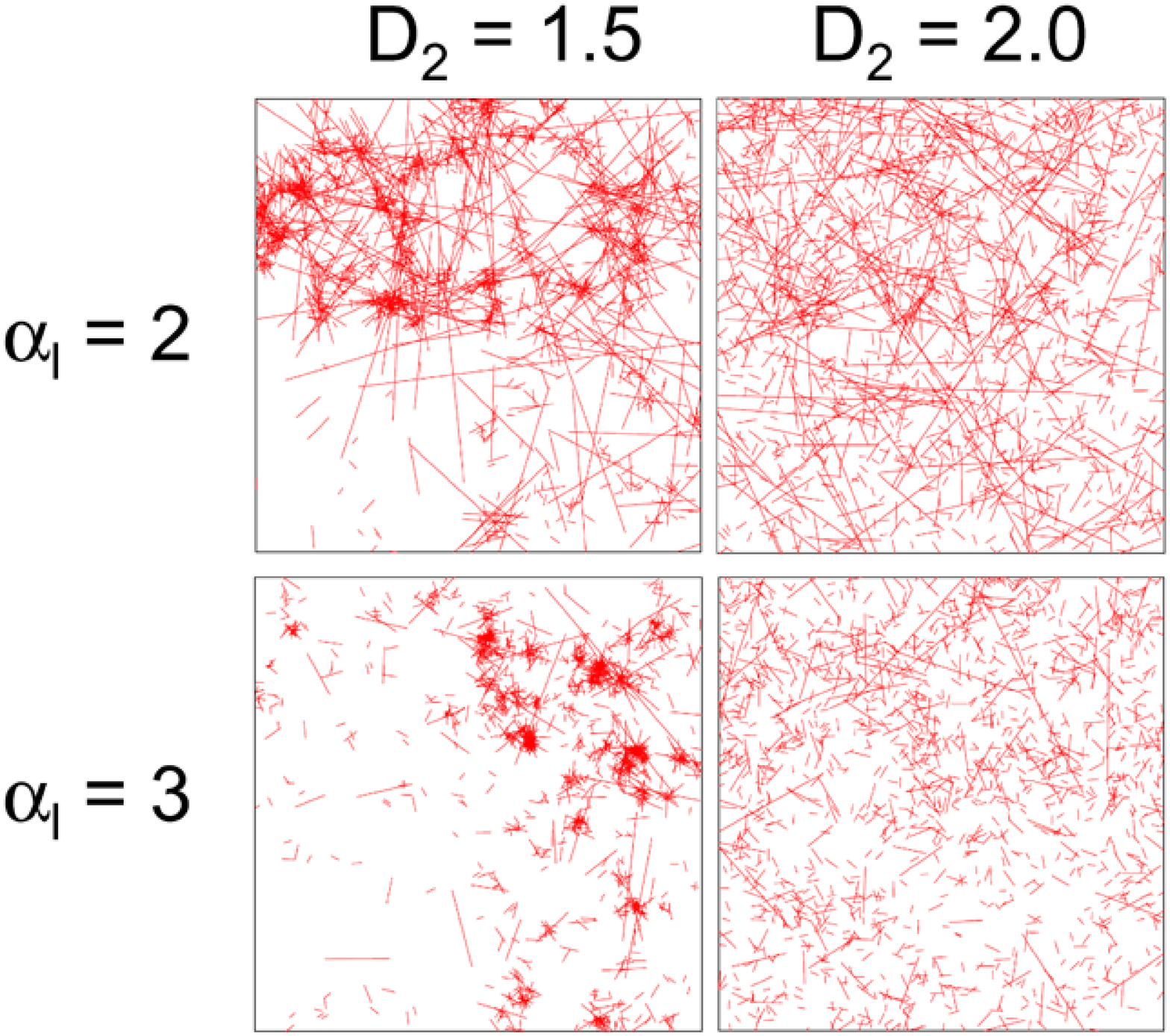}
\caption{\label{fig5} (Color online) Examples of fracture systems generated
with the DFN model for varying parameters $\alpha_l$ and $D_2$. 
} 
\end{figure}

We now turn to analyzing the DFN model \cite{dbdd03}.  It is based on 
the observation that the length of fracture lines in outcrops, $l$, are 
distributed according to a power law \cite{r99,bbod01} 
\begin{equation}
\label{pl}
p(l)\sim l^{-\alpha_l}\;.
\end{equation}
The outcrops can be divided into two groups: one with $\alpha_l=3$ 
(ASM000205 and ASM000206) and one with $\alpha_l =2.3\pm 0.2$ (the rest)
\cite{dgdbdm10}.
The angular 
distribution of the directions of the fractures depends on the fracture
system.  We assume here the simplest, i.e., a uniform distribution. The outcrop data studied show strong signs of prefered directions for the fractures, but
using this in the angular distribution of the DFN model does not have a 
significant impact.  Lastly,
the position of the fractures must be specified.  The DFN model uses a
hierarchical construction \cite{sl87,m91} to place the midpoints of the 
fractures on a fractal set characterized by a fractal dimension $D_2$. We
show in Fig.\ \ref{fig5} the resulting fracture systems for different 
parameters $\alpha_l$ and $D_2$.  The outcrop data has $D_2 \approx 2$.
Further details may be found in \cite{dgdbdm10}.

\begin{figure}
\includegraphics[scale=0.28,clip]{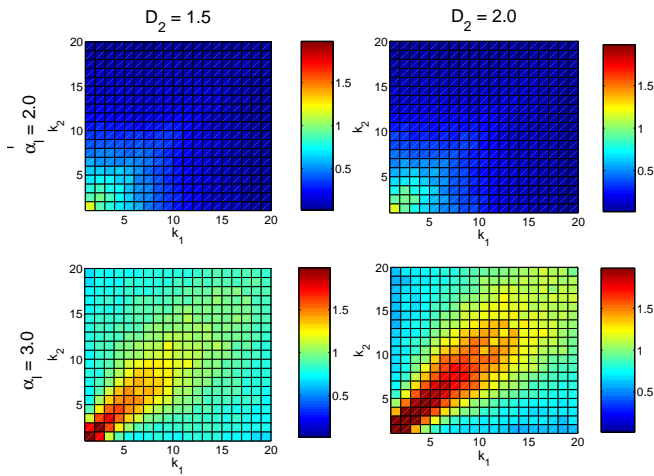}
\caption{\label{fig6} The degree correlation matrix $C(k_1,k_2)$ for
different DFN model parameters $\alpha_l$ and $D_2$.  This figure should
be compared with Fig.\ \ref{fig4}.
} 
\end{figure}

The results of analyzing the equivalent networks of the DFN model networks
are given in Table \ref{table2}. The data are based on 1000 networks of
comparable size to those in the outcrop fracture data sets.  From the table,
we see the same trends as those observed in Table \ref{table1} for the eight
outcrop fracture data sets and it is possible to find a combination of
$\alpha_l$ and $D_2$ to make match between them.  However, we show in Fig.\
\ref{fig6} the averaged degree correlation matrix.  This indicates an
{\it assortative\/} network structure: nodes of equal coordination number
tend to be connected.  This is the opposite of what is observed for the 
outcrop data sets, see Fig.\ \ref{fig4}. Hence, the topology of the
artifical networks is quite different from the natural ones.  This implies
that the topology of the fracture network themselves, artifical and real, are
quite different.  This difference is not visible from direct observation.

Hence, by constructing the equivalent networks, we have access to the entire
analysis toolbox of modern network theory for fracture networks.  As we have
shown in the analysis presented, this makes it possible to test
fracture network models on a quantitative level beyond what has been possible
earlier.    

\begin{table}
\caption{List of degree distribution power-law
exponents $\alpha_k$, clustering coefficient C,
clustering coefficient for rewired networks
$C_{RW}$, clustering coefficient for comparable random networks $C_{RA}$,
efficiency E, efficiency for rewired networks $E_{RW}$,
and efficiency  for comparable random networks $E_{RA}$
for various fracture length power-law exponents $\alpha_l$.}
\begin{tabular}{c c c c c c c c c}
\hline
\hline
$\alpha_l$ & $\alpha_k$ &  $C$  & $C_{RW}$ &
$C_{RA}$ &   $ E$  & $E_{RW}$ & $E_{RA}$\\
\hline
2.00       &   2.2     & 0.08  & 0.019   &
0.047    &  0.028  & 0.042   &  0.11  \\
2.25       &   1.7     & 0.11  & 0.013   &
0.031    &  0.027  & 0.049   &  0.11  \\
2.50       &   1.4     & 0.17  & 0.013   &
0.019    &  0.037  & 0.083   &  0.10  \\
2.75       &   1.2     & 0.26  & 0.014   &
0.014    &  0.050  & 0.134   &  0.09  \\
3.00       & 0.9/1.3\footnote{A kink in the slope around $k=60$ 
gives 0.9 when fitting for smaller values of $k$ and 1.3 for larger values.}   
& 0.31  & 0.013   &
0.008    &  0.050  & 0.154   &  0.07  \\
\hline
\hline
\end{tabular}
\label{table2}
\end{table}

We thank Svensk K\"{a}rnbr\"{a}nslehantering AB for outcrop data, H.\ F.\ Hansen and E.\ Skjetne for discussions and the referees for valuable comments. 
C.\ A.\ A.\ and A.\ H.\ thank
Statoil and The Norwegian Academy of Science and Letters for funding through 
their VISTA program. S.\ M.\ H.\ and A.\ H.\ thank the Norwegian Research
Council for funding through the CLIMIT program, grant no.\ 199970.

\end{document}